\documentclass[submit]{smj}

\Author{Touqeer Ahmad\Affil{1}, 
        Carlo Gaetan\Affil{2} 
        and Philippe Naveau\Affil{3}
}
\AuthorRunning{Touqeer Ahmad \textrm{et al.}}

\Affiliations{

\item \answer{ENSAI, CREST-UMR9194, Bruz, France.}

\item Dipartimento di Scienze Ambientali, Informatica e Statistica,
Universit\`a {Ca’ Foscari-Venezia}, Italy.

\item Laboratoire des Sciences du Climat et de l’Environnement, UMR 8212, CEA-CNRS-UVSQ, IPSL $\&$ Université Paris-Saclay, Gif-sur-Yvette, France.

}   

\CorrAddress{ENSAI, 51 Rue Blaise Pascal, 35170, Bruz, France}
\CorrEmail{touqeer.ahmad@ensai.fr}
\CorrPhone{(+33)\;750\;011\;746}
\CorrFax{(+33)\;750\;011\;746}

 \Title{
\answer{An extended generalized Pareto regression model for count data}
}
\TitleRunning{An extended generalized Pareto regression model for count data
}

\Abstract{
The statistical modeling of discrete extremes has received less attention than their continuous counterparts in the Extreme Value Theory (EVT) literature. One approach to the transition from continuous to discrete extremes is the modeling of threshold exceedances of integer random variables by the discrete version of the generalized Pareto distribution.
However, the optimal choice of thresholds defining exceedances remains a problematic issue. \answer{Moreover, in a regression framework, the treatment of the majority of non-extreme data below the selected threshold is either ignored or separated from the extremes. To tackle these issues, we expand on the concept of employing a smooth transition between the bulk and the upper tail of the distribution.} 
In the case of zero inflation, we also develop models with an additional parameter. To incorporate possible predictors, we relate the parameters to additive smoothed predictors via an appropriate link, as in the generalized additive model (GAM) framework. A penalized maximum likelihood estimation procedure is implemented. 
We illustrate our modeling proposal with a real dataset of avalanche activity in the French Alps. With the advantage of bypassing the threshold selection step, our results indicate that the proposed models are more flexible and robust than competing models, \answer{such as the negative binomial distribution}.
}

\Keywords{
Extreme value theory; discrete extended generalized Pareto distribution; zero-inflated models; generalized additive models
}

\newcommand{\answer}[1]{#1}

\newcommand{\answerr}[1]{#1}
\begin{document}

\maketitle

\section{Introduction}\label{sec:1}
Modeling count data, i.e., non-negative integers, in the presence of covariates is a very common task in many research areas.
The Poisson regression model is often of limited use because count data typically exhibit overdispersion (i.e., the variance of the counts appears larger than the mean) and/or an excessive number of zeros. Additional parameters can be inserted to deal with overdispersion, e.g., the quasi-Poisson model \citep{Wedderburn:1974}, or different distributions, such as negative binomial distribution, can be fitted. To model zero inflation, \citet{lambert1992zero} studied a two-component mixture model, one component is a point mass at zero and the other component is an assumed parametric count distribution. Lambert's specification is an example of a distributional regression model \citep{Stasinopoulos-et-al-2018,kneib-et-al:2023}. \answer{The term  ``distributional'' emphasizes \answerr{that several characteristics} of the conditional distribution of the data are modeled in terms of covariates, rather than only the mean.}

So far, the main focus of the literature has been on the relationship between the location, the scale, and the shape with the covariates \citep{rigby2005generalized}, less attention has been paid to the tail of the count distribution.

	

\begin{figure}[h]
	\begin{center}	
		\begin{tabular}{cc}
			 \includegraphics[width=0.50\linewidth]{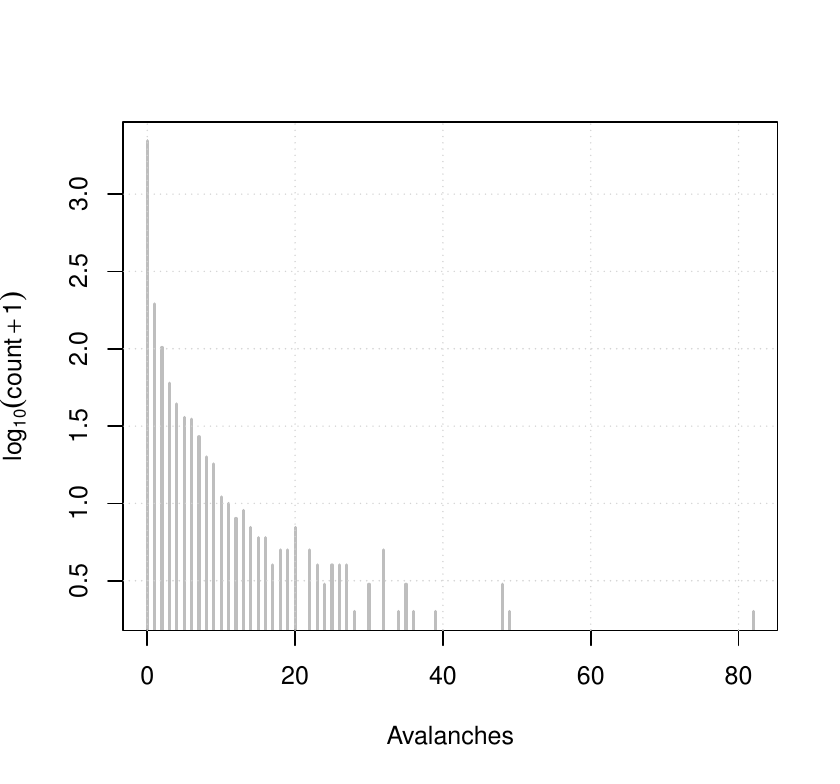}
	&
				\includegraphics[width=0.50\linewidth]{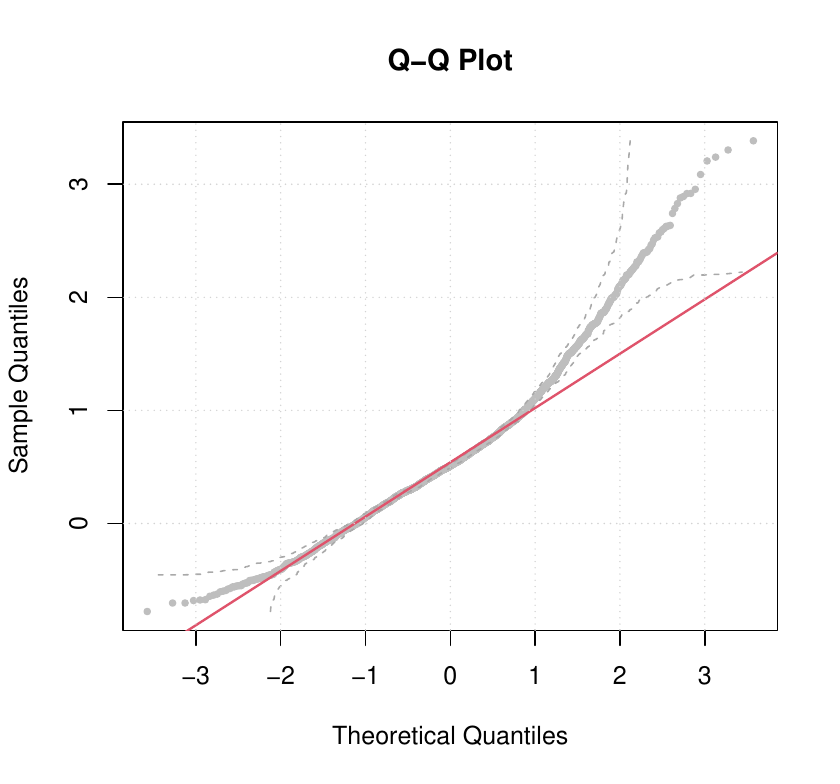}  \\
			(a) & (b) 
		\end{tabular}	
	\end{center}
	
	\caption{\answer{(a) \answerr{Frequency table plot ($\log_{10}$ scale) of avalanche events in the  Haute-Maurienne massif of the French Alps}, (b) Q-Q plot of randomized residuals from the zero-inflated negative binomial model with additive covariates. The dotted lines show the $95\%$ point-wise confidence intervals.}}
	\label{Fig:motive}
\end{figure}

\answer{As a motivating example, we consider a dataset on the avalanche activity 
in the Haute-Maurienne region of the French Alps,
The avalanche activity is measured by a three-day moving sum of the avalanche events recorded from February 1982 to April 2021,  see \citet{evin2021extreme} for details.}
 Figure \ref{Fig:motive}(a) shows a large relative frequency of zeros (meaning no avalanche has been reported)  as well as heavy-tailed behavior.

\answer{We fit a zero-inflated negative binomial regression model under the framework of generalized additive models for location, scale, and shape (GAMLSS)  where the parameters are related to additive environmental covariates (see Section \ref{sec:6} for a detailed description) via suitable link functions \citep{Stasinopoulos-et-al-2018}.  The randomized quantile residuals \citep{dunn1996randomized} are used to check the adequacy of the fitted model.
Figure \ref{Fig:motive}(b) clearly shows that the fitted models do not correctly estimate the upper tail behavior of avalanche extremes. In addition, the number of zeros is not correctly predicted in this example.}

Extreme value theory, originally developed by \citet{fisher1928limiting}, provides a mathematical blueprint to model very high and very low-frequency events (e.g., extreme temperatures, heavy rainfall intensities, heavy floods, and extreme winds, etc.), and monographs such as \citet{coles2001introduction} or \citet{beirlant2004statistics} discuss the main extreme value models.
In particular, under the \textit{peak-over-threshold} (POT) approach \citep{pickands1975statistical}, the distribution of \answerr{exceedances of a high threshold} is often approximated by the Generalized Pareto Distribution (GPD).
Modifications of GPD to discrete data exist in the literature \citep{krishna2009discrete,buddana2014discrete,kozubowski2015discrete}, and recently \citet{hitz_davis_samorodnitsky_2024} \answer{discussed} discrete versions of GPD to approximate the tail behavior of integer-valued random variables. 
This approach still requires the definition of a threshold at a high quantile, which is not easy due to the discrete nature of the data \citep{daouia:2023}. 

It should also be noted that especially environmental time series are rarely stationary and depend on environmental factors. A standard approach to modeling continuous extremes of a non-stationary process focuses on maintaining a predetermined threshold but treating parameters of the GPD as functions of \answer{covariates} \citep{davison1990models}.
An alternative approach \citep{eastoe2009modelling} uses preprocessing methods to model the non-stationarity in the body of the process to produce transformed data and then uses standard methods to model the extremes of the transformed data. 
The first approach has been adapted to the discrete case by \citet{ranjbar2020modelling}. The second approach seems to be difficult to adapt. The distribution of the preprocessed data cannot be connected to a distribution of count data.

\answer{The proposed model addresses the issue of the POT approach ignoring or separating non-extreme data below the selected threshold from the extremes. The model utilizes a smooth transition between the bulk and upper tail of the distribution, for the full range of the data, while bypassing a threshold selection.  The discrete extended version of GPD (DEGPD) is derived by discretizing the cumulative distribution function (CDF) of an extended GPD \citep{naveau2016modeling}. 
The model takes into account the possible effects of covariates in a non-parametric way.
Since it is possible to have a dataset with an excess of zeros, such as in the motivating example, we also consider a mixture of the previous distribution with a degenerate distribution at zero. This results in a distribution named Zero-Inflated DEGPD (ZIDEGPD).}

The paper is organized as follows. Section \ref{sec:2} introduces the DEGPD. The extension to deal with many zeros and covariate effects is given in Section \ref{sec:zi-reg}.  Section \ref{sec:6} discusses applications of DEGPD and ZIDEGPD to avalanche data with environmental covariates. Finally, Section \ref{sec:7} concludes with a summary of our results and a discussion of future research directions.

\section{Discrete extreme modelling}\label{sec:2}
The distribution of exceedances (i.e., the amount of data that appears over a given high threshold) is often approximated by the  Generalized Pareto distribution (GPD) defined by its  CDF as
\begin{equation} \label{eq:1}
	F(z;\sigma,\xi) =
	\begin{cases} 
		1-\left(1+\xi  z/\sigma\right)_{+}^{-1/\xi} & \xi\neq 0 \\
		1- \exp{(-{z}/{\sigma})} &  \xi = 0
	\end{cases},
\end{equation}
\answer{where $(a)_{+}= \max(a,0)$.  The $\sigma>0$ and $-\infty<\xi<+\infty$ represent the scale and shape parameters of the distribution, respectively. }

More precisely let $X$ be a random variable taking values in $[0, x_F )$ where $x_F \in(0,\infty)\cup \{\infty\} $ 
 Suppose that there exists a strictly positive sequence $a_u$ such that the distribution of 
  $a_u^{-1}  (X - u) | X \ge u$ weakly converge to a non-degenerate probability distribution on \answer{$[0, \infty)$ as $u\rightarrow x_F$}, then this distribution is the GPD \citep{balkema:de_haan:1974}.
 Thus, for large u, 
 $$\Pr (X - u > x | X \ge u) = \Pr (a_u^{-1} (X - u) >  a_u^{-1} x | X \ge u)\approx 1-	F(x;a_u\sigma,\xi).$$ 

The shape parameter, $\xi$, defines the tail behavior of the GPD. If $\xi<0$, the upper tail is bounded. \answerr{If $\xi=0$, we have the exponential distribution}, where all moments are finite. If $\xi>0$, the upper tail is \answerr{unbounded and the higher moments ultimately become infinite}. The three defined cases are labeled ``short-tailed'', ``light-tailed'', and ``heavy-tailed'', respectively. 
\answer{These categorizations enhance the flexibility of the GPD and underscore its adaptability to various modeling scenarios}.

Using the GPD  to approximate the distribution tail for discrete data can be inappropriate, as pointed out in \citet{hitz_davis_samorodnitsky_2024}.
These authors proposed to approximate the distribution tail of a count random variable $Y$ by discretizing the CDF defined by \eqref{eq:1} and, for large $u$,
\answer{
\begin{equation} \label{eq:4}
	\Pr(Y-u=k|Y\ge u) = F(k+1;\sigma,\xi)- F(k;\sigma,\xi), \hspace{1cm} k \in \mathbb{N}_{0},
\end{equation}}
with $\sigma>0$ and $\xi\ge 0$.
\answer{The distribution is called discrete GPD (DGPD), and several properties of discrete Pareto type distributions have been studied previously in the literature \citep{krishna2009discrete,buddana2014discrete,kozubowski2015discrete}.}


A drawback of GPD in the continuous case is that it only models observations that occur above a certain high threshold.
This imposes an artificial dichotomy in the data (i.e., observations are either below or above the threshold), and finding the optimal threshold remains complex for practitioners. 
The choice becomes more complicated when the observations feature a substantial number of ties.  

In the continuous extreme value setting, many authors have attempted to model the full range of data without threshold selection \citep{frigessi2002dynamic,carreau2009hybrid,macdonald2011flexible,papastathopoulos2013extended,naveau2016modeling,stein:2021}. 

Notably, \citet{papastathopoulos2013extended} proposed an extension of GPD that incorporated an additional shape parameter without affecting the tail behavior. The inclusion of this parameter stabilized the GPD parameter estimates for threshold selection, allowing a lower threshold to be selected. In order \answer{to work with the} modeling of the lower tail and the bulk of the distribution, \citet{naveau2016modeling} identified two conditions that ensured compliance with the EVT for the lower and upper tails. The transition between the two tails was modeled by a function on $[0,1]$, which can take different forms.
For example, \citet{Tencaliec-et-al:2020} worked with a Bernstein polynomial base. \citet{decarvalho2022} extended this type of approach to Bayesian lasso structures. The main ingredient of all these constructions, called \answer{extended generalized Pareto distributions} (EGPDs), is the idea of the integral transformation to simulate GPD random draws, i.e. $F_{\sigma,\xi}^{-1}( U),$ where $U\sim\mathcal{U}(0,1)$ represents a uniformly distributed random variable on $(0,1)$ and $F_{\sigma,\xi}^{-1}$ denotes the inverse of the CDF \eqref{eq:1}. This leads to the family of distribution for the random variable 
	\begin{equation}\label{eq:3}
		Z=F_{\sigma,\xi}^{-1}\left(G^{-1}(U) \right),
	\end{equation}
	where $G$ is a CDF on $[0,1]$ and $U\sim\mathcal{U}(0,1)$. Clearly, the CDF of $Z$ is $G(F(z;\sigma,\xi))$. The key problem is to find a function $G$ which preserves the upper tail behavior with shape parameter $\xi$ and also controls the lower tail behavior. \citet{naveau2016modeling} defined   restrictions for validity of $G$ functions. For instance, the tail of $G$  denoted  by $\Bar{G}=1-G$ has to satisfy 
\begin{eqnarray}
	\lim_{u \to 0} \frac{\Bar{G}(1-u)}{u}&=&a,  \mbox{ for some finite $a>0$ (upper tail behavior),}\label{eq:cond1}\\
	\lim_{u \to 0} \frac{G(u)}{u^\kappa}&=&c,  \mbox{ for some finite $c>0$ (lower tail behavior).}\label{eq:cond2}
\end{eqnarray}
Four parametric examples for  $G$ have been proposed in \citet{naveau2016modeling} (see also below).
\answerr{We follow} the same idea and define the probability mass function (pmf) for the count variable $Y$ as
\answer{
\begin{equation}\label{eq:DEGPD}
	\Pr(Y=y) = G\left(F\left({y+1};\sigma,\xi\right)  \right)- G\left(F\left({y};\sigma,\xi\right)  \right),\qquad  y\in \mathbb{N}_0.
\end{equation}}
The distribution defined by (\ref{eq:DEGPD}) is referred to as the discrete extended generalized Pareto distribution (DEGPD).
 The explicit formula of  CDF of DEGPD is developed as
 \begin{equation}\label{eq:8}
 	\Pr(Y\le y)= G\left(F\left({y+1};\sigma,\xi\right)\right)
 \end{equation} 
 and the quantile function is derived as
 \begin{equation}\label{eq:9}
 	q_{p}=
 \left\{
 	\begin{array}{ll}
 		\left\lceil{\frac{\sigma}{\xi}\left\{ \left(1- G^{-1}(p) \right)^{-\xi}-1\right\}}\right\rceil-1, & \mbox{if } \xi>0\\
   \\
 		\left\lceil{-\sigma \log \left(1- G^{-1}(p) \right)}\right\rceil -1,& \mbox{if } \xi=0\\
 	\end{array}
 \right.
 \end{equation} 
 with $0<p<1$.  In this paper,  we use  four parametric expressions of $G(\cdot)$
\citep{naveau2016modeling}, namely
\begin{description}
	\item[Model (i):] \label{c1} 
	$G(u;\psi)=u^{\kappa}$, $\psi=\kappa >0$;
	\item[Model (ii):]\label{c3}
	$G(u;\psi)=1-D_{\delta}\{(1-u)^{\delta}\}$, $\psi=\delta>0$ where $D_{\delta}$ is the CDF of a Beta random variable with parameters $1/\delta$ and $2$, that is:
	$$D_{\delta}(u)=\frac{1+\delta}{\delta} u^{1/\delta} \Big( 1-\frac{u}{1+\delta} \Big);$$
	\item[Model (iii):] \label{c4}
	$G(u;\psi)=[1-D_{\delta}\{(1-u)^{\delta}\}]^{\kappa/2}$, $\psi=(\delta, \kappa)$ with $\delta>0$  and $\kappa>0$;
	\item[Model (iv):] \label{c2}
	$G(u;\psi)=p u^{\kappa_1}+(1-p)u^{\kappa_2}$, $\psi=(p,\kappa_{1},\kappa_{2})$ with $\kappa_2 \geq \kappa_1 >0$ and $p \in (0,1)$.
	
\end{description}

\begin{figure}
\begin{center}
\begin{tabular}{ccc}
	\includegraphics[width=0.45\linewidth]{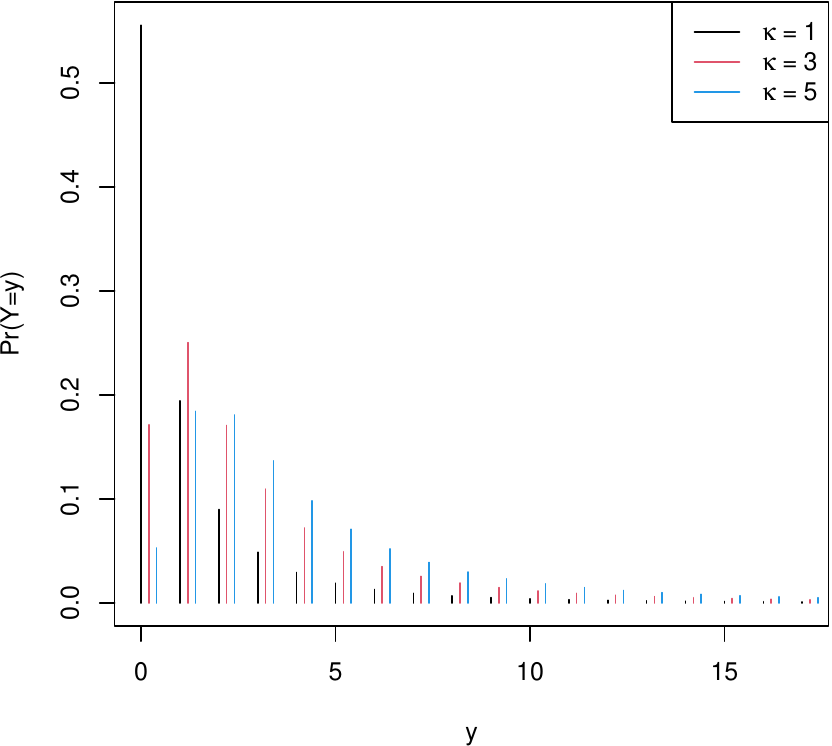}  
&\hspace{1cm}&
	\includegraphics[width=0.45\linewidth]{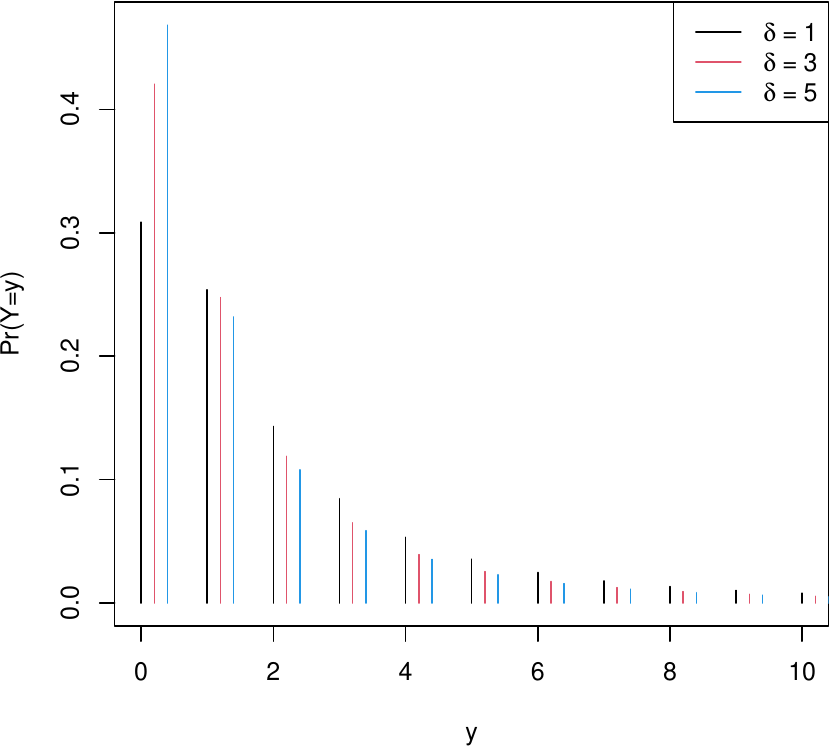}  
\\
(a) && (b)
\\	
\vspace{0.3cm}\\
\includegraphics[width=0.45\linewidth]{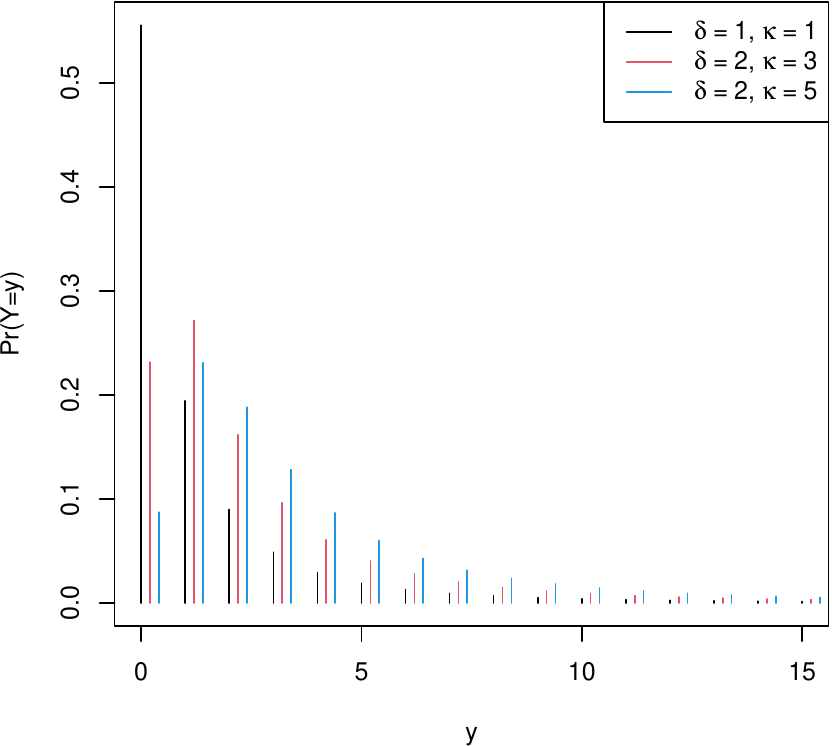}  
&&
\includegraphics[width=0.45\linewidth]{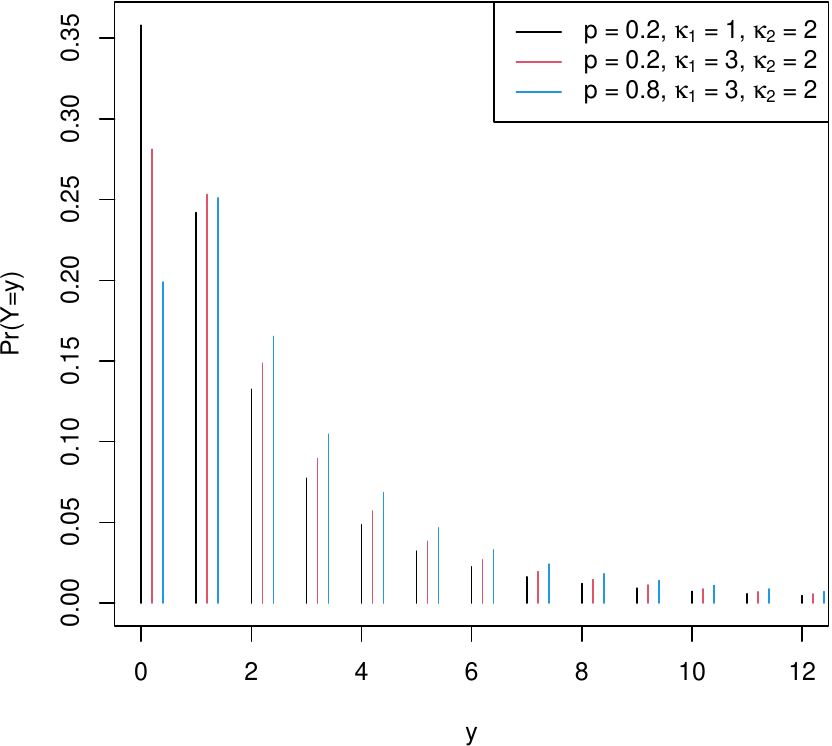}  
\\
(c) && (d)
\end{tabular}	
\end{center}	
	\caption{Probability mass function corresponding to model  \eqref{eq:DEGPD} for $\sigma=1$, $\xi=0.5$. (a)   Model (i): $G(u;\psi)= u^{\kappa}$,   $\kappa=1,3,5$; (b) Model (ii): $G(u;\psi)= 1- D_{\delta}(1-u)^{\delta}$  $\delta=1,3,5$; (c) Model (iii): $G(u;\psi)= [1- D_{\delta}(1-u)^{\delta}]^{\kappa/2}$ for 
		 $\delta= 1, 2,2$ and $\kappa=1, 3, 5$, and (d) Model (iv): $G(u;\psi)= p u^{\kappa_1}+(1-p)u^{\kappa_2}$ for $p=0.2,0.2,0.8$,  $\kappa_1=1, 3,3$ and $\kappa_2= 2, 2,2 $.}
	\label{fig:pmf}
\end{figure}

The parametric family (i)  leads to a pmf of DEGPD  with three parameters $(\kappa, \sigma$ and $\xi)$: $\kappa$ deals the shape of the lower tail, $\sigma$ is a scale parameter, and $\xi$ controls the rate of
upper tail decay. Thus, Figure \ref{fig:pmf}(a) shows the behavior \answer{of the pmf} of DEGPD with fixed scale and upper tail shape parameter (i.e., $\sigma=1$ and $\xi=0.5$) and with different values of lower tail behaviors ($\kappa$=1, 2, 5). The DGPD is recovered when $\kappa=1$, and additional flexibility for low values is attained by varying $\kappa$. For instance, more flexibility on the lower tail can be observed without losing upper tail behavior in Figure \ref{fig:pmf}(a) \answerr{when $\kappa=5$}.

The parametric family (ii) is another interesting choice for constructing DEGPDs. This choice is more complex than the previous two. Figure \ref{fig:pmf}(b) illustrates the behavior of pmf with different values of $\delta$. \answer{In the continuous setting, we observe that the EGPD associated with this G family converges to GPD as $\delta$ increases to infinity}. Furthermore, the conditions \eqref{eq:cond1} and \eqref{eq:cond2} are also satisfied with $\delta=2$,  see \citet{naveau2016modeling} for more details. In discrete settings, the DEGPD corresponding $G(u;\psi)=1-D_{\delta}\{(1-u)^{\delta}\}$ also become \answer{very close} to the pmf of DGPD  when $\delta$ increases to infinity. In general, the parameter $\delta$ describes the central part of the distribution. Thus, this parameter relatively improves the flexibility for the central part of the distribution. The $\delta$ parameter can be interpreted as a threshold tuning parameter. One of the drawbacks of DEGPD (ii) is that it models only the central and upper parts of the distribution. On the other hand, the behavior of the lower tail could not be estimated directly. 


The parametric family (iii) supports the lower tail of the distribution with the $\kappa>0$ parameter. Interestingly, this family also tends to the DEGPD with parameters ($\kappa,\delta,  \sigma$ and $\xi$). The $(\kappa,\delta$ and $\xi)$ represents the lower, central, and upper parts of the distribution, respectively, and $\sigma$ is a scale parameter as usual. In particular, Figure \ref{fig:pmf}(c) shows the behavior of pmf of  DEGPD linked with $G(u;\psi)=\left[1-D_{\delta}\{(1-u)^{\delta}\}\right]^{\kappa/2}$ at different settings of the parameters. \answer{Changing the values of $\kappa$ also reveals \answerr{flexibility of the lower tail.}}

The parametric family (iv) is the mixture of power laws: $\kappa_1$ identifies the lower tail shape, $\kappa_2$ modifies the central distribution shape, and $\sigma$ and $\xi$ are scale and upper tail parameters, respectively. It can be observed from Figure \ref{fig:pmf}(d) that the DEGPD related to $G(u;\psi)=p u^{\kappa_1}+(1-p)u^{\kappa_2}$ is also showing flexibility with \answer{\mbox{$p=0.2, 0.8$}}, $\sigma=1$, $\xi=0.5$, $\kappa_1=1, 3$, constant values of $\kappa_2=2$.

\section{Zero-inflation and regression modeling}\label{sec:zi-reg}

As we have seen for our motivating example, many zeros can be found in various real data sets. In that case, the current model with a flexible lower and upper tail cannot be adjusted for the excessive zeros. We follow \citet{lambert1992zero} and change DEGPD's pmf to
	\begin{equation}\label{eq:ZIDEGPD}
		\Pr(Y=y)=\left\{
		\begin{array}{ll}
			\pi +(1-\pi)G(F\left(1,\sigma,\xi\right);\psi ) & y=0 \\
			(1-\pi)\left[G\left(F(y+1,\sigma,\xi);\psi \right) -G\left(F(y,\sigma,\xi);\psi) \right)
			\right] & y\in \mathbb{N}
		\end{array}
		\right.
	\end{equation}
	where $0\leq \pi \leq1$ is the mixing proportion, determining from which state $Y$ is generated. In the following, we coin  \eqref{eq:ZIDEGPD} as the ZIDEGPD model.

	Suppose now that $\boldsymbol{x}\in \mathbb{R}^q$ is a vector of covariates measured with $Y$.  
	 In a continuous framework, the inclusion of flexible forms of dependence of extreme values on covariates is well established \citep{davison1990models}.
	 \citet{chavez2005generalized} used the Generalized Additive Model (GAM) \citep{hastie1990generalized} to flexibly estimate GPD parameters.  More recently, by coupling GAM forms with penalized likelihood, 	 \citet{youngman2019generalized} modeled threshold exceedances with GPD parameters of GAM forms.
	
By allowing the parameters to depend on covariates, we extend the pmf \eqref{eq:ZIDEGPD} to the zero-inflated count regression setting. More specifically, we identify the vector of parameters $(\xi,\sigma,\psi,\pi)$ with $\theta = (\theta_1, ..., \theta_d)$. 
The parameters of the distribution of $Y$ can depend on the covariates $\boldsymbol{x}$, i.e. $\theta(\boldsymbol{x})=(\theta_1(\boldsymbol{x}),\ldots,\theta_d(\boldsymbol{x}))$. 	To relate the distribution parameters $(\theta_1(\boldsymbol{x}),\ldots,\theta_d(\boldsymbol{x}))$ to the covariates, we consider additive predictors of the form
\answer{\begin{equation}\label{eq:predictors}
		\eta_i(\boldsymbol{x})=s_{i1}(\boldsymbol{x})+\cdots+s_{iJ_i}(\boldsymbol{x}),\qquad i=1,\ldots,d,
	\end{equation}}
	where $s_{i1}(\cdot),\ldots, s_{iJ_i}(\cdot)$ are smooth functions of the covariates $\boldsymbol{x}$. The predictors are linked to the parameters via known monotonic and twice differentiable link functions $h_i(\cdot)$.
	\begin{equation}\label{eq:link}
		\theta_i(\boldsymbol{x})=h_i(\eta_i(\boldsymbol{x})), \qquad i=1,\ldots, d.
	\end{equation}
	\answer{For instance, we use the following linking functions for the model (i) associated with $G(u;\psi)=u^\kappa$}. \answerr{The parameters can be written as}
	\begin{align*}
\xi(\boldsymbol{x})=\exp(\eta_\xi(\boldsymbol{x})),\,\, \sigma(\boldsymbol{x})=\exp(\eta_\sigma(\boldsymbol{x})),\,\, 
\kappa(\boldsymbol{x})=\exp(\eta_{\kappa}(\boldsymbol{x})),\,\,
	\pi(\boldsymbol{x})=
	\exp\left(\frac{\eta_{\pi}(\boldsymbol{x})}{1+\eta_{\pi}(\boldsymbol{x})}\right).	    
	\end{align*}

\answer{The functions $s_{ij}(\cdot)$ in \eqref{eq:predictors} are approximated by a set of $K_{ij}$  basis functions $\{B_{k,ij}(\boldsymbol{x})\, k=1,\ldots, K_{ij}\}$, namely }
	\begin{equation}\label{eq:basis}
		s_{ij}(\boldsymbol{x})=\sum_{k=1}^{K_{ij}} \beta_{ij,k}B_k(\boldsymbol{x}).
	\end{equation}
 The basis functions can be of different types \citep[see][for instance]{wood2017generalized}. The basis function expansions can be written as 
	$
	s_{ij}(\boldsymbol{x})=  \boldsymbol{t}_{ij}(\boldsymbol{x})^T\boldsymbol{\beta}_{ij}
	$
	where $\boldsymbol{t}_{ij}(\boldsymbol{x})$ is still a vector of transformed covariates that depends on the basis functions
	and $\boldsymbol{\beta}_{ij}=(\beta_{ij,1},\ldots, \beta_{ij,K_{ij}})^T$ is a parameter vector to be estimated.
	
	The penalized maximum likelihood estimation (MLE) method is used to estimate the parameters of the proposed models. More precisely, let $y_1, \ldots, y_n$ be $n$ independent observations from \eqref{eq:DEGPD} and $\boldsymbol{x}_1,\ldots,\boldsymbol{x}_n$ the related covariates. The log-likelihood function is given by
	\begin{eqnarray}\label{eq:likzi}
		l(\boldsymbol{\beta})&=& \sum_{i=1}^n I_{\{0\}}(y_i)\log \left[\pi(\boldsymbol{x}_i)+(1-\pi(\boldsymbol{x}_i))
		G(F(1;\sigma(\boldsymbol{x}_i), \xi(\boldsymbol{x}_i));\psi(\boldsymbol{x}_i))\right] \nonumber \\ 
		&&+\sum_{i=1}^{n}(1-I_{\{0\}}(y_i)) \log(1-\pi(\boldsymbol{x}_i)))\times \nonumber\\
		&& \left[ G(F(y_{i}+1;\sigma(\boldsymbol{x}_i), \xi(\boldsymbol{x}_i));\psi(\boldsymbol{x}_i))- 
		G(F(y_{i};\sigma(\boldsymbol{x}_i), \xi(\boldsymbol{x}_i));\psi(\boldsymbol{x}_i)) \right],
	\end{eqnarray}
	\answer{where $I_{A}(\cdot)$ is the indicator function of the set $A$}. To ensure regularization of the functions $s_{ij}(\boldsymbol{x})$  so-called penalty terms are added to the objective log-likelihood function. 
	Usually, the penalty for each function $s_{ij}(\boldsymbol{x})$ \answerr{is a quadratic} penalty $ \boldsymbol{\beta}_{ij}^T \boldsymbol{P}_{ij}(\boldsymbol{\lambda}_{ij}) \boldsymbol{\beta}_{ij} $ where $\boldsymbol{P}_{ij}(\boldsymbol{\lambda}_{ij})$ is a known semi-definite matrix and the vector  $\boldsymbol{\lambda}_{ij}$ regulates the amount of smoothing
	needed for the fit. A special case that we use in the real data application is when $\boldsymbol{P}_{ij}(\boldsymbol{\lambda}_{ij})=\lambda_{ij}\boldsymbol{P}_{ij}$, \answer{for a scalar $\lambda_{ij} >0$ and a semi-definite matrix $\boldsymbol{P}_{ij}$. 
 The entries of the penalty matrix $P_{ij}$ are the integrals of the products of the second derivatives of pairs of cubic spline functions, see \citet[Section 5.3]{wood:2011} for more details.}
		The penalized log-likelihood function for the latter models reads:
  \answer{
	\begin{equation}\label{eq:penlik}
		l_p(\boldsymbol{\beta}) = l(\boldsymbol{\beta}) - \frac{1}{2}\sum_{i=1}^d \sum_{j=1}^{J_i} {\lambda}_{ij}\boldsymbol{\beta}_{ij}^T \boldsymbol{P}_{ij} \boldsymbol{\beta}_{ij}.
	\end{equation}
 }
	
	We apply the restricted maximum likelihood (REML) approach to estimate $\boldsymbol{\beta}_{ij}$ and $\boldsymbol{\lambda}_{ij}$ following \citet{wood:2011}. \answer{Our current implementation exploits the R  package \texttt{evgam} \citep{youngman2020evgam} and adds two new families of distributions named \texttt{degpd} and \texttt{zidegpd}. 
Note that within the   \texttt{degpd} family, \answerr{all four} models for $G(\dot,\psi)$ have been implemented. \answerr{By contrast}, for the \texttt{zidegpd} family, only models (i), (ii), and (iii) have been implemented, since the \texttt{evgam} package allows the simultaneous estimation of only five distribution parameters $\theta(\boldsymbol{x})=(\theta_1(\boldsymbol{x}),\ldots,\theta_5(\boldsymbol{x}))^T$.
%
}

\section{Avalanche  data example}\label{sec:6}
Let us return to the example introduced with Figure \ref{Fig:motive}. The data comes from
the \textit{Enquête Permanente sur les Avalanches}.
This survey collects avalanche data from the French Alps and has monitored about 3900 routes by local observers since the beginning of the 20th century \citep[see][]{mougin1922avalanches,evin2021extreme}. Quantitative (run-out elevations,
deposit volumes, etc.) and qualitative (flow regime, snow quality, etc.) information is collected for each event. 

\answer{Report quality of avalanches observed by local observers varies over time and space. For example, an avalanche event may be recorded a few days after it occurs, and the estimated day of the event by the observer may be approximate. It is too restrictive to select only the events for which the day is known, as too many events would be lost. However, if the avalanche occurred several days before the observation, including the observation could lead to biased analysis, such as an inaccurate count of avalanches for that day and difficulty in linking the event to snow and meteorological conditions \citep{evin2021extreme}. Therefore, we only consider avalanche events that occurred \answerr{within three days before} of their observation.} 

Natural avalanche activity is also uncertain because records tend to record paths visible from valleys, so the high-elevation activity may be underestimated. Avalanches are usually caused by severe storms that bring high snowfalls coupled with snow drifting, but substantial variations of environmental covariates causing snow melt and/or fluctuations of the freezing point can also be involved. 
For instance, precipitation amounts, air temperatures during storms, and prior snow stratigraphy influence avalanche types and frequency. 
Although overall avalanche frequency is likely to decrease globally \citep{strapazzon2021effects}, more extreme environmental conditions during winter storms can cause more intensive avalanche events. For instance, a shallow snow-pack and warmer temperatures have become increasingly influential in the Alps.
Since extreme events have potentially terrible consequences, it is crucial to anticipate future avalanche activity in the short-term and long-term management, possibly relating this activity to environmental variables.

\answer{
In our example the total number of avalanches in the Haute Maurienne massif was considered within a three-day observation window between January 1982 and April 2021. Only days between October 15th and May 15th were included.
A total of 2839 observations were made during this period \citet{dkengne2016limiting}.}

 The related environmental variables are maximum wind speed (WS)   at 10 meters in m/s,
relative humidity (RH) at 2 meters in percentage,
precipitation (PREC) in mm per day, maximum (MxT) and minimum (MnT) temperatures at 2 meters in $^o$C.   
 Environmental covariates have been downloaded from  \href{https://power.larc.nasa.gov/data-access-viewer/}{https://power.larc.nasa.gov/data-access-viewer/} by specifying latitude and longitude information.  \answer{Three-day summaries are calculated by averaging RH and PREC and taking the maximum for MxT and WS and the minimum for MnT.}

\begin{figure}[H]
	\begin{center}	
	\begin{tabular}{ccc}
 		\includegraphics[width=0.32\linewidth]{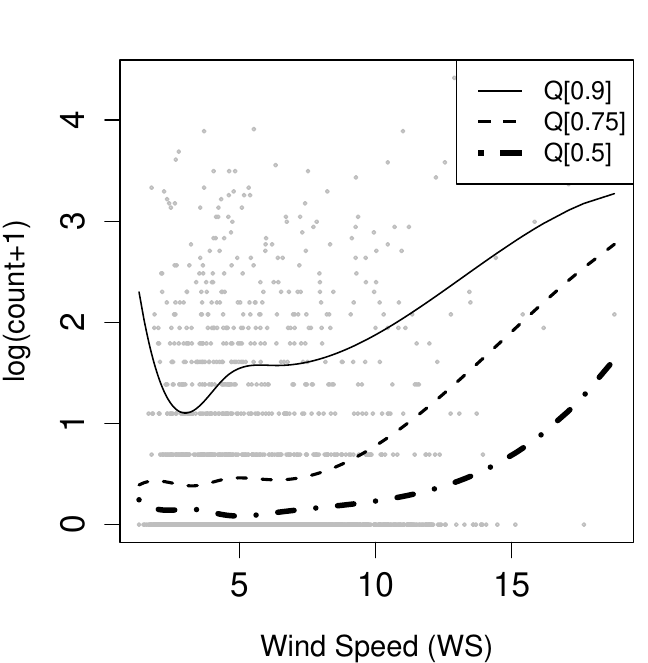}
		&
		\includegraphics[width=0.32\linewidth]{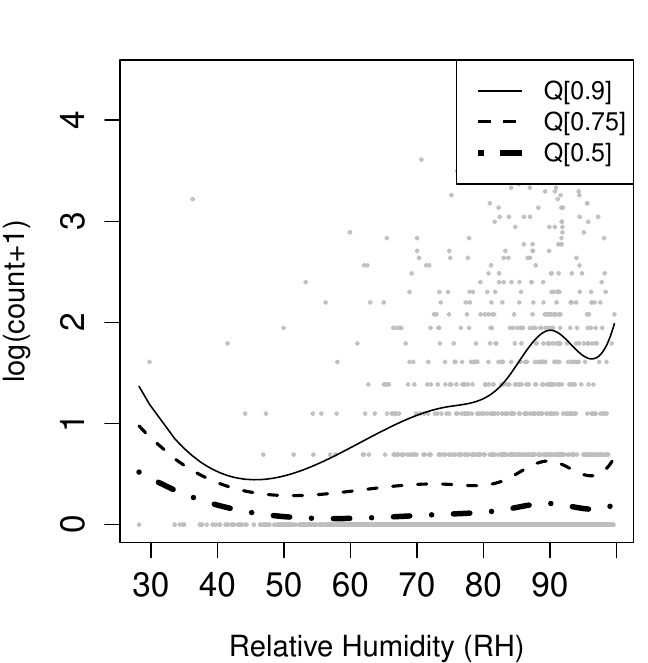}
		&\includegraphics[width=0.32\linewidth]{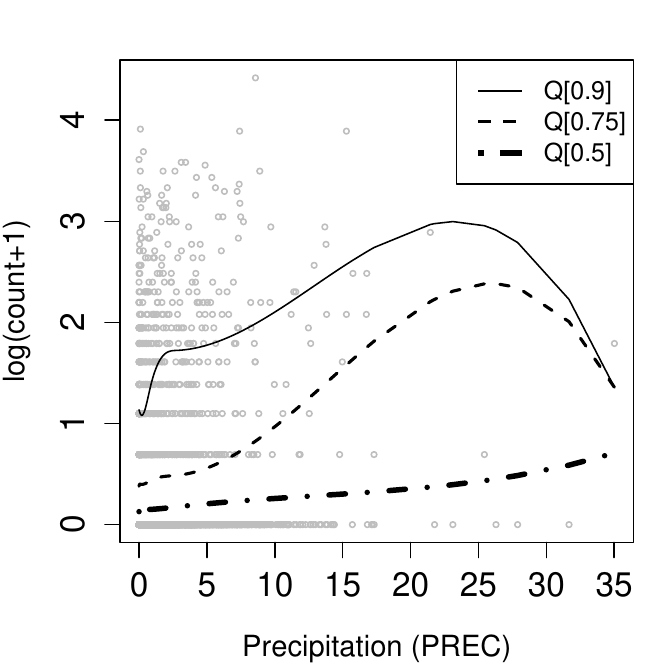}
   \\
  \includegraphics[width=0.32\linewidth]{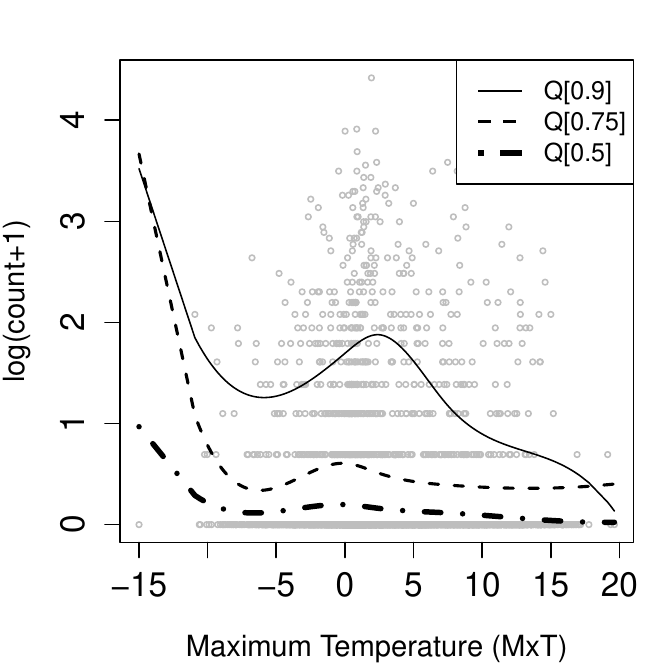}
		&
		\includegraphics[width=0.32\linewidth]{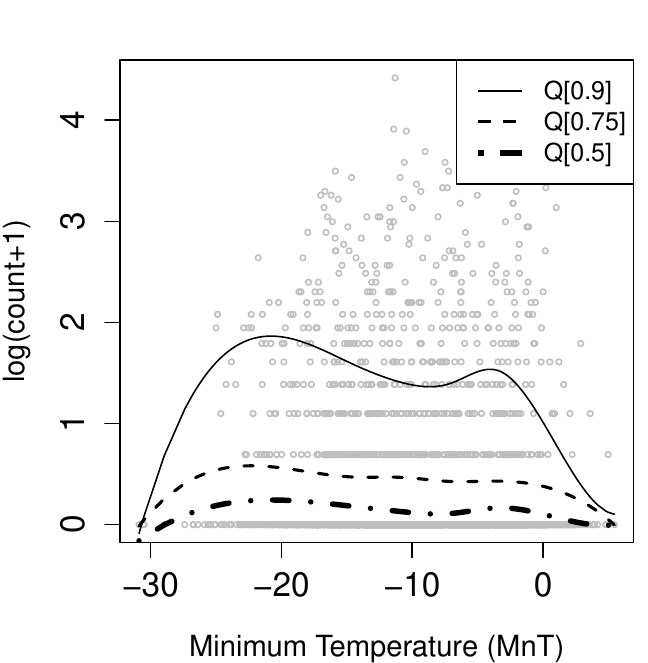}
	\end{tabular}	

	\end{center}
	\caption{\answer{Scatterplots of avalanche counts (log scale) versus environmental variables.  The smoothing lines} are obtained by jittering the count and fitting three non-parametric regression quantile models at the 0.5, 0.75, and 0.90 quantiles.}
	\label{fig:scatt}
\end{figure} 

		  

\begin{figure}[t]
	\centering
	\includegraphics[width=0.7\linewidth]{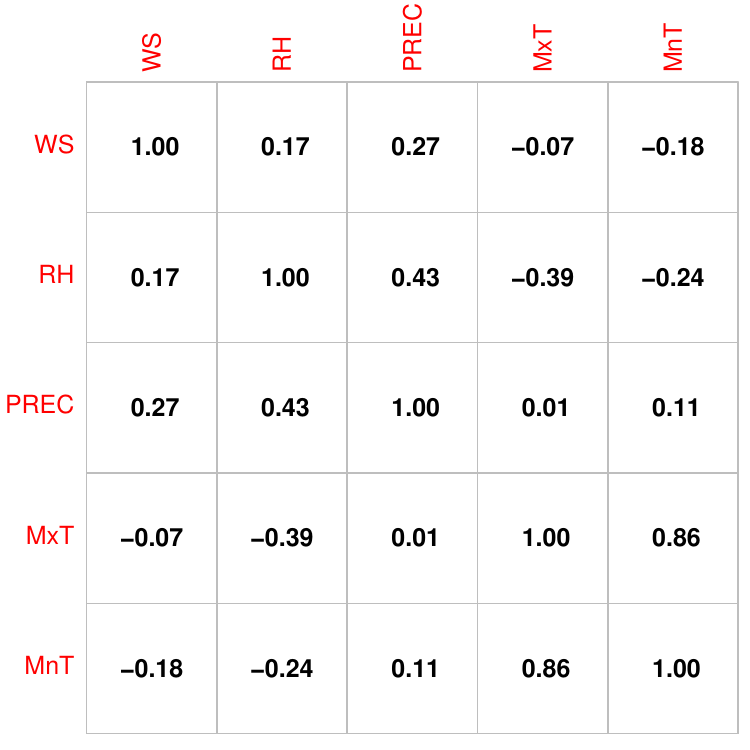}  
	\caption{Correlation between environmental variables.}
	\label{fig:corr}
\end{figure}

\answer{The scatterplots of avalanche counts (log scale)} versus environmental variables (Figure \ref{fig:scatt}) do not highlight clear relationships, mainly masked by the presence of many zeros.
Figure {\ref{fig:corr}} shows the correlation plot between the covariates, highlighting that maximum temperature (MxT) and minimum temperature (MnT) are positively strongly correlated. At the same time, precipitation (PREC) has no significant correlation with MxT. On the other hand, relative humidity (RH) has a low and moderate positive correlation with wind speed (WS) and PREC, while it has a weak negative correlation with temperature variables. Furthermore, wind speed and precipitation have a weak correlation with minimum and maximum temperature variables.  

We fit the ZIDEGPD model under specifications (i), (ii), and (iii) for $G(u;\psi)$,
and a backward variable selection procedure based on AIC is performed for selecting the covariates.  The shape parameter $\xi$ is kept constant to avoid numerical instability in the estimation of $\psi$. The results of the selection are given in Table {\ref{tab:selection}}, where $s(\cdot)$ indicates the smoothed predictor.

\begin{table}[t]
	\caption{Selected predictors for ZIDEGPD model $^a$}\label{tab:selection}
\begin{center}
	\begin{tabular}{cllcl}
		\hline
		Model
		 &\multicolumn{1}{c}{ $\log\sigma(\boldsymbol{x})$} & \multicolumn{1}{c}{ $\log\kappa(\boldsymbol{x})$} &  $\log\delta(\boldsymbol{x})$ &
		  \multicolumn{1}{c}{$\operatorname{logit}(\pi(\boldsymbol{x}))$}\\
		  \hline
		(i)
		& \texttt{s(WS)+s(MxT)+s(PREC)} &\texttt{s(RH)}&-&\texttt{s(MxT)}\\
		(ii) 
		& \texttt{s(WS)+s(MxT)+s(RH)}& - & \texttt{cst} & \texttt{s(MxT) + s(PREC)}\\
		(iii) 
		& \texttt{s(WS)+s(MxT)+s(RH)+s(PREC)}&\texttt{cst} & \texttt{cst} & \texttt{s(MxT)}\\
		\hline
	\end{tabular}
\end{center}
    \footnotesize{$^a$ \texttt{cst} denotes the constant parameters.}
\end{table}

%

To assess the overall adequacy of ZIDEGPD model, we   used  the  randomized
residuals \citep{dunn1996randomized} defined as
\begin{equation*}
	r_i= \Phi^{-1} \left((1-u_i)F(y_i-1;\hat{\boldsymbol{\theta}}(\boldsymbol{x}_i))+u_i F(y_i; \hat{\boldsymbol{\theta}}(\boldsymbol{x}_i)  )\right)
\end{equation*}
where $\Phi^{-1}$ is the inverse of CDF of a standard Gaussian distribution function, $u_i$ is drawn from a uniform distribution, and $F(.; \boldsymbol{\mathbb{\theta}})$ is the CDF of the current model. 
Aside from sampling variation in the parameter estimates, randomized residuals should follow a standard Gaussian distribution if the model is correctly identified.
\begin{figure}[t]
	\begin{center}	
	\begin{tabular}{ccc}
 		\includegraphics[width=0.3\linewidth]{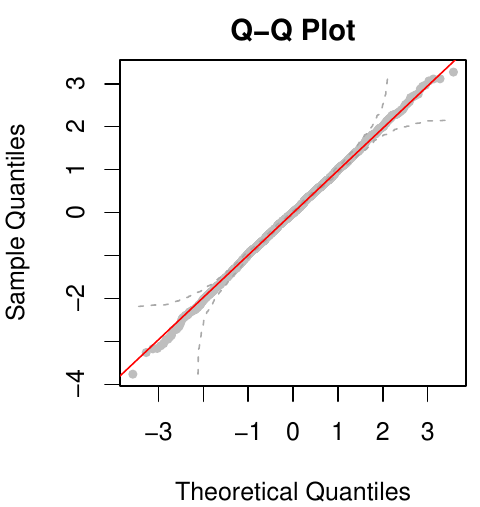}  
		&
		\includegraphics[width=0.3\linewidth]{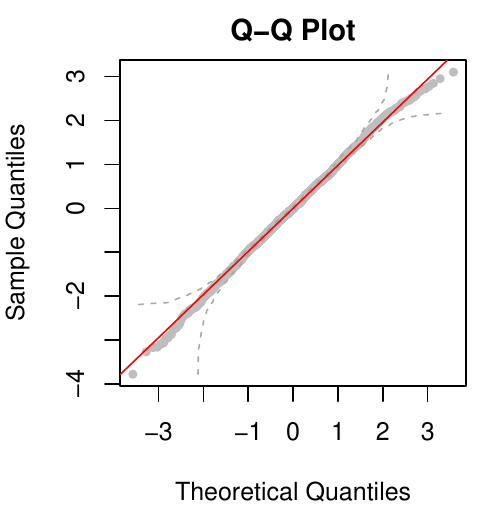}  
		&\includegraphics[width=0.3\linewidth]{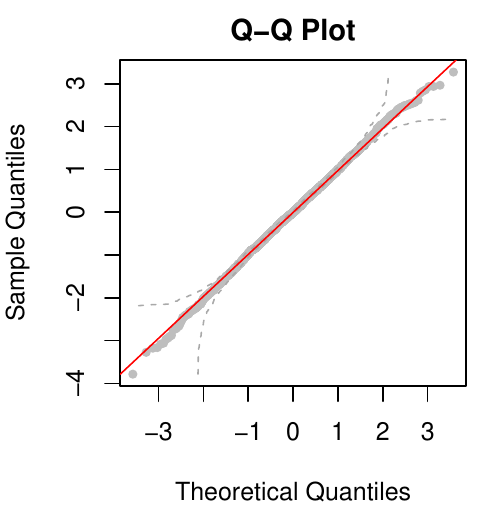}  \\
		(a) & (b) & (c)
	\end{tabular}	

	\end{center}
	\caption{Q-Q plot of randomized residuals of different ZIDEGPD models: (a) Model (i); (b) Model (ii); (c) Model (iii). \answer{The dotted lines show the $95\%$ point-wise confidence intervals.}}
 \label{fitqqplot}
\end{figure} 
The randomized residuals derived from our proposed model show no apparent departure from normality (Figure \ref{fitqqplot}) and fit both tails (lower and upper) correctly when compared to the existing models in the literature, as shown in Figure \ref{Fig:motive}. 

In Table \ref{tab:ad}, we report the final values of the AIC and values of the Anderson-Darling test for the normality of the randomized residuals.
Since looking at just one realization can be misleading, we compute the Anderson-Darling test statistic for 1,000 realizations of the randomized residuals, take the median of the test statistic, and evaluate the corresponding P-value. As we can see, Model (i) is slightly preferred in terms of goodness of fit, and Table \ref{tab:final} summarizes the output of the fitting procedure for this model. Note that the smoothing terms are all statistically significant, in particular, there is no indication for  $\kappa(x)=1$, i.e. when the DEGPD reduces to a DGPD in the mixture.
\begin{table}[]
	\centering
	\caption{AIC values and Anderson-Darling  normality test statistics 
		of the different ZIDEGPD models.}
\vspace{0.5cm}
\answer{
	\begin{tabular}{cccc}
	\hline
	\multicolumn{1}{c}{Model}&AIC& Statistics &P-value \\
	\hline
	\text{ (i)}&\answerr{6121}&  0.287& 0.623\\
	\text{ (ii)}& \answerr{6151}&0.367 &0.432\\
	\text{ (iii)}&\answerr{6152}& 0.315& 0.543\\
	\hline
\end{tabular}
}
	\label{tab:ad}
\end{table}

\begin{figure}[H] 
	\begin{center}
		\begin{tabular}{ccc}
			\includegraphics[width=0.30\linewidth]{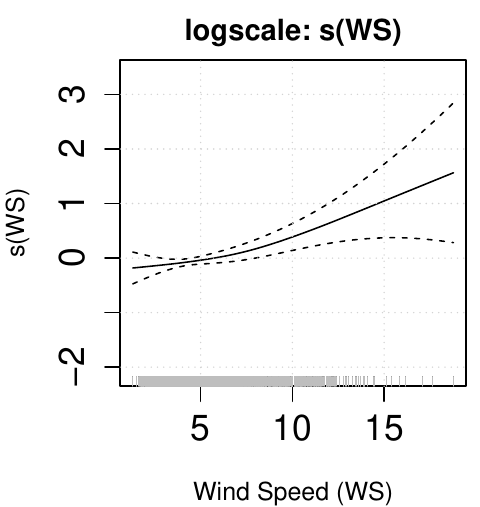}
			&
			\includegraphics[width=0.31\linewidth]{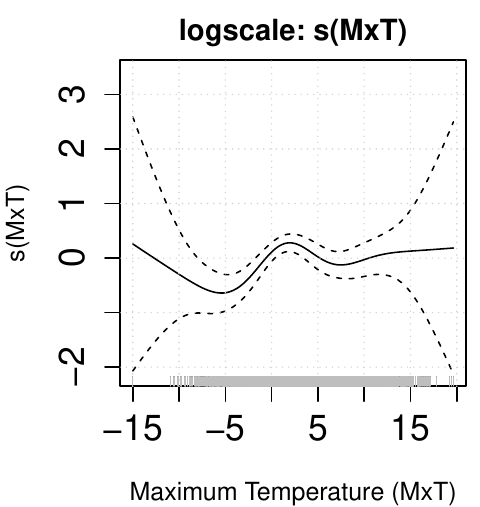}
			&	\includegraphics[width=0.31\linewidth]{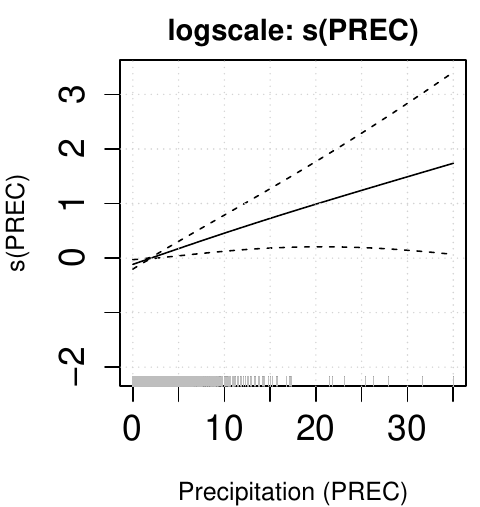}
			\\
			\includegraphics[width=0.31\linewidth]{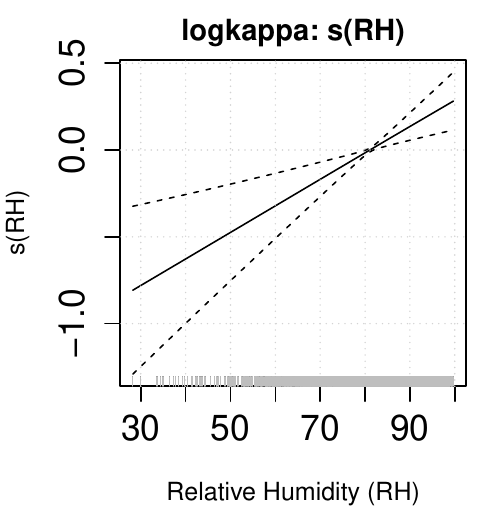}
			&
			\includegraphics[width=0.31\linewidth]{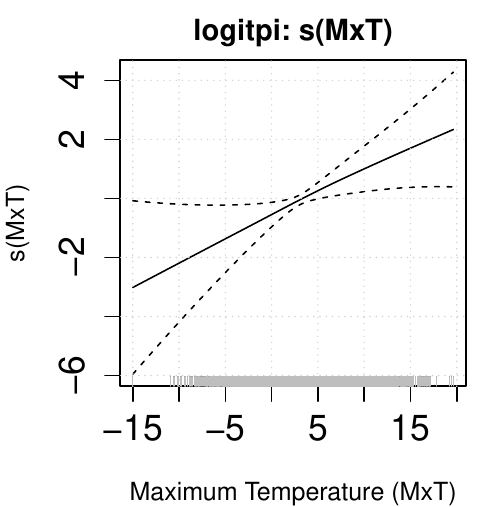}
			&			
			\\		
		\end{tabular}	
	\end{center}
	\caption{Estimated non-parametric effects of covariates in parameters of ZIDEGPD (i). 
 }	
	\label{fig:additive}
\end{figure}

\begin{table}
	\centering
	\caption{Estimated coefficients and smooth terms for  ZIDEGPD model (i) fitted to avalanches data. \answer{The p-values of the smoothed terms \texttt{s($\cdot$)} indicate the significance of their presence.}}
\vspace{0.5cm} 
\answer{
	\begin{tabular}{lrrrrrr}
		\hline		
		\multicolumn{6}{c}{\textbf{ZIDEGPD} with $G(u;\psi)= u^\kappa$}\\
		\hline
		&\multicolumn{5}{c}{{Constant terms}}\\
		\cline{2-6}
		& \multicolumn{1}{c}{Estimate} & \multicolumn{1}{c}{Std Error} & \multicolumn{1}{c}{t value}& \multicolumn{1}{c}{P-value}& \\
		\cline{2-6}
		$\log (\kappa)$ & -1.62   &    0.22  & -7.42& $<$0.001
  \\
		$\log(\sigma)$ & 1.62   &    0.15  & 10.71  & 
  $<$0.001
  \\
		$\log(\xi)$ & -1.63    &   0.43  & -3.83& $<$0.001
  \\
		$\operatorname{logit}(\pi)$ & -1.25   &    0.73  & -1.71 &  0.044
  \\
		\hline
		&\multicolumn{5}{c}{Smooth terms for $\log(\kappa)$}\\
		\cline{2-6}
		& {edf} & {max.df} & {Chi.sq} & {Pr($>|t|$)}& {}\\
		\hline
		\texttt{s(RH)} &   1.02   &   9 & 11.78 & $<$0.001
  \\
		\hline
		&\multicolumn{5}{c}{Smooth terms for $\log(\sigma)$}\\
		\cline{2-6}
		& {edf} & {max.df} & {Chi.sq} & {Pr($>|t|$)}& {}\\
		\cline{2-6}
		\texttt{s(WS)} & 1.88   &   7  &14.80 & 0.005 
  \\
		\texttt{s(MxT)} & 4.81  &    7 & 26.76 & $<$0.001
  \\
		\texttt{s(PREC)} & 1.10  &    4  & 7.16 & 0.008 
  \\
		\hline
		&\multicolumn{5}{c}{Smooth terms for $\operatorname{logit}(\pi)$}\\
		\cline{2-6}
		& {edf} & {max.df} & {Chi.sq} & {Pr($>|t|$)}& {}\\
		\cline{2-6}
    		\texttt{s(MxT)} & 1.16   &   7 &  7.32  & 0.013 
      \\
		\hline
		\noindent\end{tabular}
}
	\label{tab:final}
\end{table}
\answer{A broad interpretation of the results (see also Figure  \ref{fig:additive}) is that meteorological conditions have an increasing effect on avalanche occurrence, except for the maximum of temperature, which shows a 
non-monotonic pattern around 0 degrees. Only the temperature affects the mixing proportion $\pi$, with an increment for temperatures above $0^o$C.
The value of $\kappa$ increases with relative humidity, which is associated (see Figure \ref{fig:pmf}-(a)) with an increasing probability of extreme avalanches.} 
Our results are in agreement with those of the existing literature \answer{\citep{dreier2013influence}} where, for example, snow surface, air temperatures, and changes in snow height and relative humidity strongly influenced snow slides in spring periods. 
 
\section{Conclusions}\label{sec:7}

There are many examples of distributions for count data. Motivated by the avalanche real data example, we attempt to properly model the lower limit of the distribution and the upper tail without neglecting the bulk part of the data.

The family of distributions deriving \answerr{from an efficient choice} of the function $G(u;\psi)$ in the DEGPD offers great flexibility in their behavior in both tails.   
\answer{Additionally, we apply mixture models within this framework to further enhance modeling capabilities, particularly in addressing excess zeros commonly observed in such datasets.}

We have developed and implemented R functions to fit DEGPD and ZIDEGPD parameters in GAM forms that allow for non-identically distributed discrete extremes. These functions use the functions in \texttt{evgam} \citep{youngman2020evgam}.
\answer{As a result, it is possible to consider non-additive model formulations and fit them using thin-plate splines, which are particularly attractive for modeling isotropic spatial dependence, or tensor products of splines for modeling interactions between covariates.}

In terms of application, the variability of avalanche activity has been statistically related to other environmental variables (e.g., temperature, wind, precipitation, and humidity).
 \answer{Although this model does not allow for short-term predictions, it does allow for the association of specific weather conditions with avalanche risk by allowing experts to account for possible nonlinearity.
 Indeed the \answerr{GAMs} proposed in this study allow parametric and non-parametric} functional forms, which would most likely be required for larger data sets. Compared to other competing models available in the literature our proposed models are more flexible in estimating both tail behavior \answer{and achieving} a better fit for avalanche data when environmental conditions are considered as covariates.

\section*{Acknowledgements}
 The authors thank Benjamin Youngman for helpful discussions on the R code and Nicolas Eckert for providing and explaining the avalanche data. Touqeer Ahmad acknowledges support from the Région Bretagne through project SAD-2021-
MaEVa. Philippe Naveau acknowledges the support of the French Agence Nationale de la Recherche (ANR) under reference ANR-Melody (ANR-19-CE46-0011). 
Part of this work was also supported by 80 PRIME CNRS-INSU, ANR-20-CE40-0025-01 (T-REX project), and the European H2020 XAIDA (Grant agreement ID: 101003469).

\section*{Code availability}
Full source code for the proposed models with a simple running example is available at  
\href{https://github.com/touqeerahmadunipd/degpd-and-zidegpd}{https://github.com/touqeerahmadunipd/degpd-and-zidegpd}\,.

\bibliography{smj-template}

\end{document}